%% file: proceeding.tex
\documentclass[10pt]{article}
\usepackage{graphicx}

\usepackage[bookmarksopen=true,bookmarksnumbered=true]{hyperref}%

\def\Title#1{\begin{center} {\Large #1 } \end{center}}
\def\Author#1{\begin{center}{ \sc #1} \end{center}}
\def\Address#1{\begin{center}{ \it #1} \end{center}}

\newcommand\pubblock{\rightline{\begin{tabular}{l} Proceedings of the Fifth Annual LHCP\\ \pubnumber\\
         \pubdate  \end{tabular}}}

\newenvironment{Abstract}{\begin{quotation} \begin{center} 
             \large ABSTRACT \end{center}\bigskip 
      \begin{center}\begin{large}}{\end{large}\end{center} \end{quotation}}

\newenvironment{Presented}{\begin{quotation} \begin{center} 
             PRESENTED AT\end{center}\bigskip 
      \begin{center}\begin{large}}{\end{large}\end{center} \end{quotation}}

\input econfmacros.tex

\usepackage{color}
\usepackage{lineno}
\usepackage{float}

\newcommand\pubnumber{ ATL-PHYS-PROC-2017-099 }

\newcommand\pubdate{\today}

\def\affiliation{
On behalf of the ATLAS and CMS Collaborations, \\
INFN, Sezione di Bologna}

\usepackage[subrefformat=parens,labelformat=parens]{subfig}
\usepackage{placeins}

\begin{document}

\large
\begin{titlepage}
\pubblock

\vfill
\Title{Boosted top production in ATLAS and CMS}
\vfill

\Author{ Marino Romano  }
\Address{\affiliation}
\vfill
\begin{Abstract}
An overview of the boosted top production analyses using data collected by the ATLAS and CMS experiments at $\sqrt{s}=$ 8 TeV and 13 TeV of proton-proton
collisions at the LHC is presented. These analyses use techniques for the reconstruction of boosted objects to measure the production of top quarks at high transverse
momenta.  The measurements are optimized for the different final states and
for different ranges of the transverse momenta of the particles involved, improving on  measurements with traditional objects reconstruction based on the combination of resolved objects. 
\end{Abstract}
\vfil

\begin{Presented}
The Fifth Annual Conference\\
 on Large Hadron Collider Physics \\
Shanghai Jiao Tong University, Shanghai, China\\ 
May 15-20, 2017
\end{Presented}
\vfill
\end{titlepage}
\def\thefootnote{\fnsymbol{footnote}}
\setcounter{footnote}{0}
%

\normalsize 


\input{introduction}

\input{reconstruction_short}

\input{cross_section_short}

\input{charge_asymmetry_short}

\input{summary}

\end{document}

%% file: econfmacros.tex
\def\beq{\begin{equation}}
\def\eeq#1{\label{#1}\end{equation}}
\def\eeqn{\end{equation}}

\def\beqa{\begin{eqnarray}}
\def\eeqa#1{\label{#1}\end{eqnarray}}
\def\eeqan{\end{eqnarray}}

\let\bar=\overbar

\def\Dslash{\not{\hbox{\kern-4pt $D$}}}
\def\dslash{\not{\hbox{\kern-2pt $\del$}}}

\def\msb{{\bar{\ssstyle M \kern -1pt S}}}

\newcommand{\tev}{\,\mathrm{TeV}}
\newcommand{\gev}{\,\mathrm{GeV}}

\newcommand{\pb}{\,\mathrm{pb}}
\newcommand{\fb}{\,\mathrm{fb}}

\newcommand{\invfb}{\,\mathrm{fb}^{-1}}
\newcommand{\ttbar}{{t\bar{t}}}
\newcommand{\pt}{p_{\mathrm{T}}}
\newcommand{\Fig}{Figure~\ref}
\newcommand{\Tab}{Table~\ref}

\newcommand{\virg}{``}
\newcommand{\thad}{t_\mathrm{had}}

%% file: introduction.tex
\section{Introduction}

The top quark is the heaviest quark, significantly heavier than its partner, the bottom quark. Once the bottom quark was experimentally discovered in 1977, the existence of a charge $2/3$ quark in the third quark generation was expected to preserve the Standard Model (SM) renormalizability. 
The top quark was discovered by both the CDF \cite{top:cdf} and D$\emptyset$ \cite{top:d0} collaborations in 1995 at the Tevatron collider.

The top quark is special not only due to its large mass, but also due to its short lifetime which prevents it from hadronizing before decaying, i.e. there are no bound state hadrons made of top quarks. This allows to experimentally test the properties of the \virg bare'' top quark itself through its decay products without diluting information in the hadronization process.

As {the properties of the top quark} are precisely predicted by the SM, top quark physics provides sensitive probes to test the validity of the SM and a tool to investigate the Higgs boson properties and to potentially discover physics beyond the SM.

Top quark pairs, $\ttbar$, are produced in proton-proton collisions via the strong interaction of the colliding partons. In a leading order picture,  at $\sqrt{s}=8$ TeV  about 85\% of $\ttbar$ pairs are  produced  by gluon-gluon interactions. The
remaining 15\% is produced in quark and anti-quark annihilation processes.

 A top quark decays almost exclusively into a $W$ boson and $b$ quark. Therefore, $\ttbar$ events can be classified through the decay products of the two $W$ bosons. $W$ bosons can decay into a quark pair or a lepton-neutrino pair, so it is possible to define three final state topologies: full hadronic, when both $W$ bosons decay in a quark pair; dileptonic, when both $W$ bosons decay in a lepton-neutrino pair; lepton+jets, when a $W$ boson decays in a quark pair and the other in a lepton-neutrino pair. 

When the top quark is produced with a very high momentum, its decay products are
collimated, affecting the reconstruction efficiency of each individual decay product.
This specific phase space is referred to as ``boosted regime'', and can be particularly interesting when probing the Standard Model at extreme energies or when searching for heavy particles decaying in pair of top quarks. In
addition, a boosted topology study can complement a resolved analysis by
recovering events otherwise mis-reconstructed, or can provide additional advantages,
such as a reduction of the combinatorial background due to less final state objects.

%% file: reconstruction_short.tex
\section{Boosted top reconstruction in ATLAS and CMS}

The defining property of boosted object decays is the fact that their decay products
 appear collimated in the momentum direction of the boosted mother particle in the
 rest frame of the detector. Therefore, the decay products merge into a single large
 radius (large-$R$) jet with a characteristic substructure that can be exploited to distinguish
 these jets from those initiated from gluons or light-quarks.

 At high-luminosity hadron
 colliders, a major obstacle for analyses relying on large-$R$ jets is the presence of pile-up
 and the underlying event, both of which lead to soft, wide-angle contaminations that
 dilute the jet substructure. Their impact is  mitigated using the so-called \virg grooming'' techniques. A
 large number of combinations of large-$R$ jet reconstruction algorithms, substructure
 variables and grooming algorithms to identify boosted top-quarks have
 been studied in the ATLAS \cite{ATLAS} and CMS \cite{CMS} experiments. \Tab{tab:boosted_techniques}
 summarizes the jet reconstruction and top tagging
 tools generally used by each experiment. 

\begin{figure}[H]
\centering
\subfloat[\label{fig:ATLAS:tag:eff:80}]
{
\includegraphics[width=0.45\linewidth]{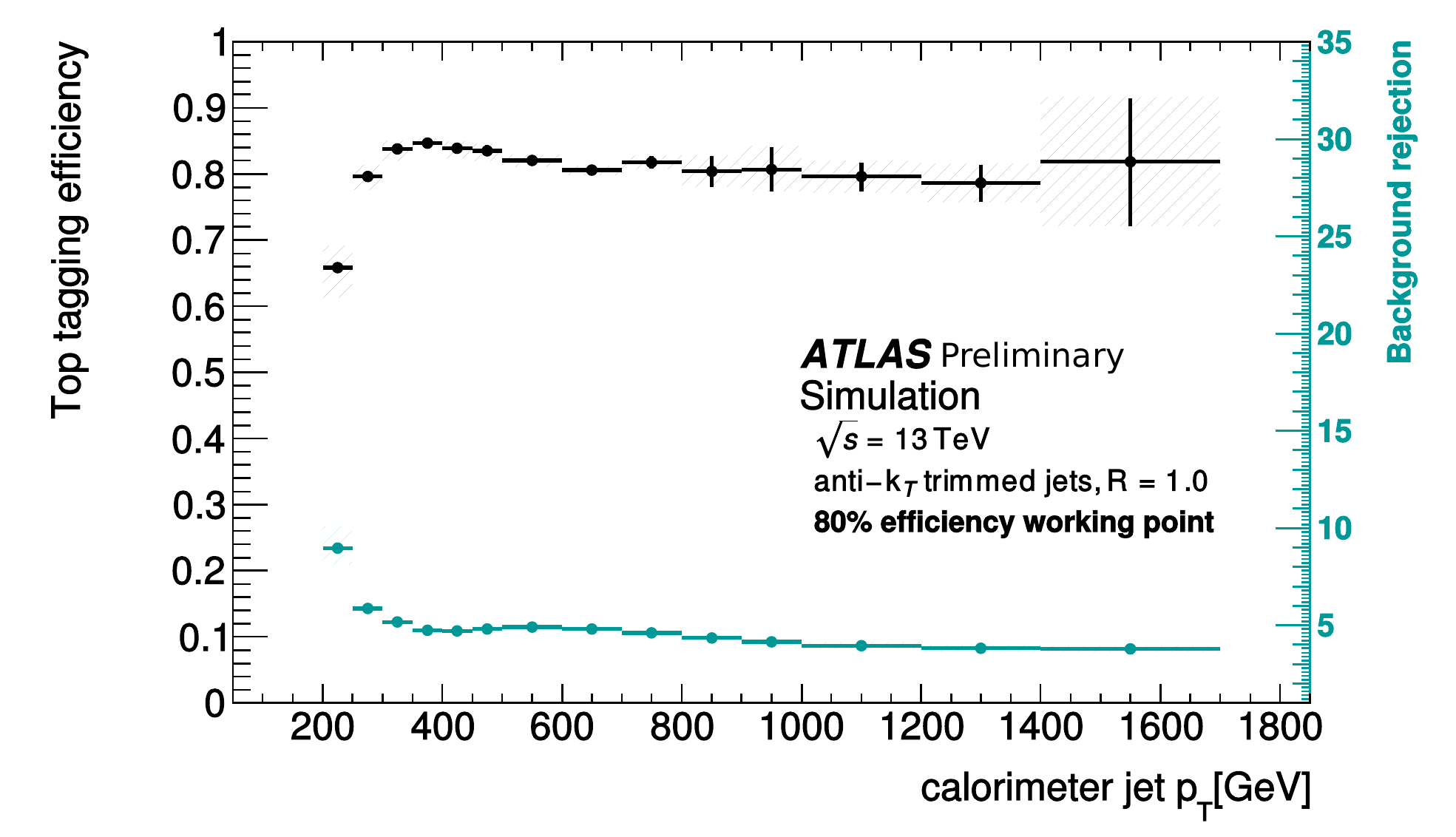}
}
\subfloat[\label{fig:CMS:tag:eff}]
{
\includegraphics[width=0.33\linewidth]{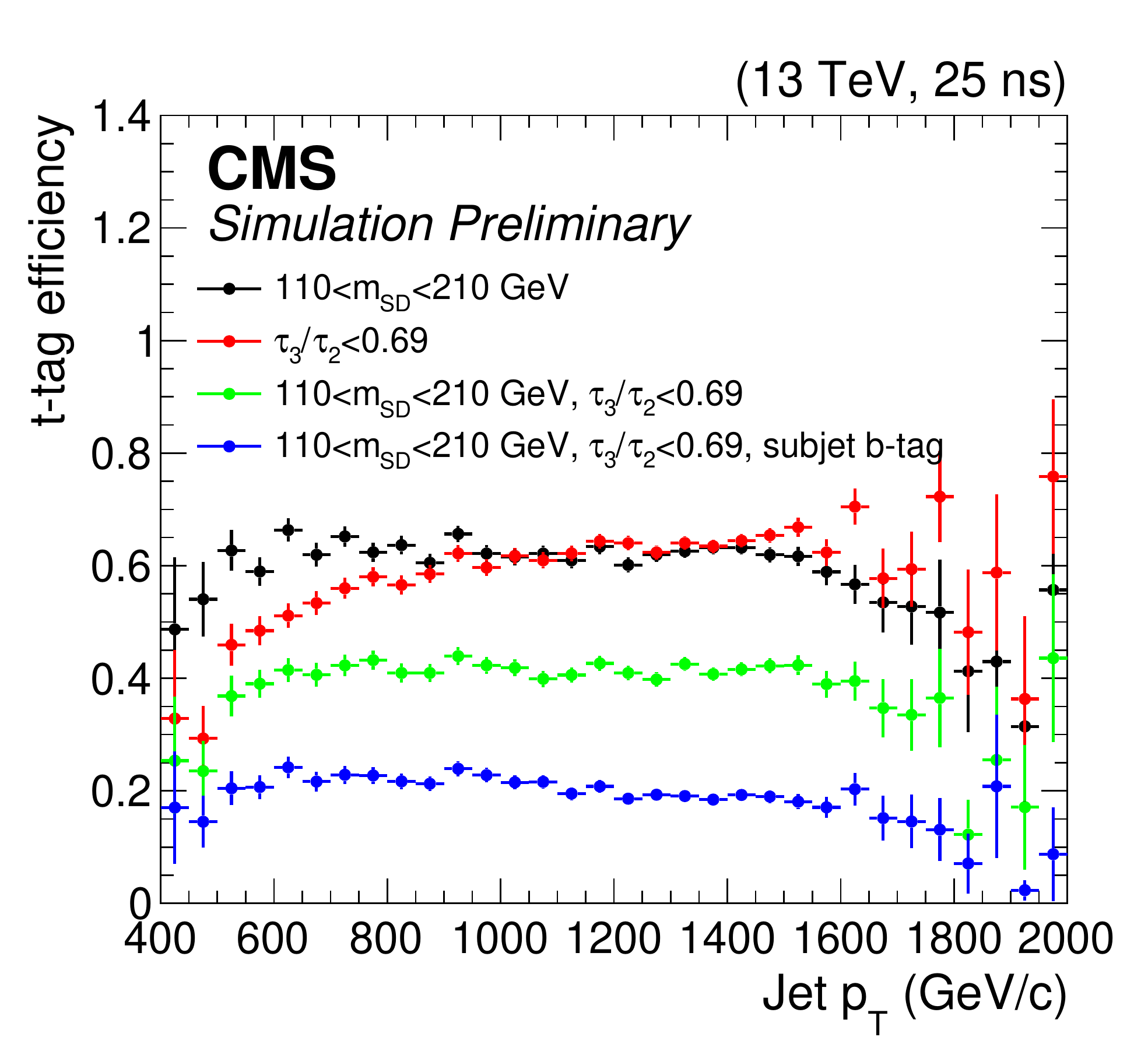}
}
\caption{\protect\subref{fig:ATLAS:tag:eff:80} top-tagging efficiencies \cite{ATLAS:boosted:reco:13tev} as a function of the $\pt$ of the reconstructed jet for the \virg$80\%$'' working points used in the differential cross section measurements performed by ATLAS in the lepton+jets  channel \cite{ATLAS:xs:ljets:13tev}.  \protect\subref{fig:CMS:tag:eff} top-tagging efficiencies as a function of the $\pt$ of the reconstructed jet for different top-tagging criteria employed by the CMS experiment \cite{CMS:boosted:reco:13tev}.\label{fig:tag:eff}}
\end{figure}

The rapid growth of the field makes it
 impossible to cover every single technique within the scope of this document. The full
 extent of those studies is documented in \cite{ATLAS:boosted:reco:8tev,ATLAS:boosted:reco:13tev} for ATLAS and \cite{CMS:boosted:reco:8tev,CMS:boosted:reco:13tev} for
 CMS. The top-tagging efficiencies as a function of the  transverse momentum of the reconstructed jets for  several working points used by ATLAS and CMS are shown in \Fig{fig:tag:eff}.

\begin{table}[H]
\begin{center}
\begin{tabular}{|c|c|c|}  
\hline & ATLAS & CMS\\
\hline Jet reconstruction & anti-$k_T$ $R=1.0$ & C/A $R=1.5$ and anti-$k_T$ $R=0.8$\\
\hline Grooming & Trimming, $R_\mathrm{subjet}=0.2$, $f_\mathrm{cut}<5\%$ & Soft-drop, $z_\mathrm{cut}=0.1$ and $\beta=0$\\
\hline Top tagging & Jet mass, $d_{12}$ (8 TeV only) and $\tau_{3/2}$ & Jet mass, $\tau_{3/1}$ and $\tau_{3/2}$\\
\hline 
\end{tabular}
\caption{Summary of the jet reconstruction, grooming and top tagging  criteria employed by ATLAS and CMS.
\label{tab:boosted_techniques}}
\end{center}
\end{table}
%
%
%
%
%

%% file: cross_section_short.tex
\section{Top pair differential cross section measurements}
Since the predictions for the inclusive $\ttbar$  production cross section are in very good agreement with the measurements  and considering the rapid increase of the integrated luminosity, the focus is starting to switch to the ``differential'' measurements of the top properties, such as the cross section  as a function of the $\ttbar$ system kinematic variables. Such experimental measurements, performed in different channels, allow precision tests of the predictions of perturbative QCD. New physics may also give rise to additional $\ttbar$  production mechanisms or modifications of the top quark decay channels, that can be discovered looking at the differential spectra. 

Top pair differential cross section in the boosted regime has been measured in the lepton+jets and full hadronic channels by ATLAS \cite{ATLAS:xs:ljets:13tev,ATLAS:xs:ljets:8tev,ATLAS:xs:fullhad:13tev}, and CMS \cite{CMS:xs:ljets:8tev,CMS:xs:fullhad:13tev,CMS:xs:jetmass} at a center of mass energy of 8 and 13 TeV.

\subsection{Top pair differential cross sections in the lepton+jets channel
at 13 TeV}
The measurement of $t\bar{t}$ differential cross sections in the lepton+jets decay channel has been performed by ATLAS, in both the resolved and boosted regimes, at
a center-of-mass energy of $\sqrt{s}=13\tev$ \cite{ATLAS:xs:ljets:13tev}. The integrated luminosity corresponds to $L = 3.2 \invfb$,
recorded in 2015 in proton-proton collisions. In the boosted regime, the absolute ($\frac{d\sigma}{dX}$) and relative ($\frac{1}{\sigma}\frac{d\sigma}{dX}$) differential cross sections have been measured as a function of the transverse momentum and absolute rapidity of the hadronically decaying top ($X=\pt^{\thad},\,\left|y^{\thad}\right|$).

 Boosted top quark events in the lepton+jets channel are selected by requiring exactly one charged lepton (electron or muon), at least one small-$R$ jet
and exactly one large-$R$ jet. The large-$R$ jet is tagged as a top using the working point corresponding to an efficiency of $80\%$ of the tagger described in \cite{ATLAS:boosted:reco:13tev}. The measurement is performed at particle level:  the reconstructed distributions are corrected to the particle-level by means of a bayesian-inspired unfolding technique \cite{unfolding:bayes}.
 
\begin{figure}[htbp]
\centering

\subfloat[\label{fig:ATLAS:diffxs:pt:ljets:13tev}]
{
\includegraphics[width=0.32\linewidth]{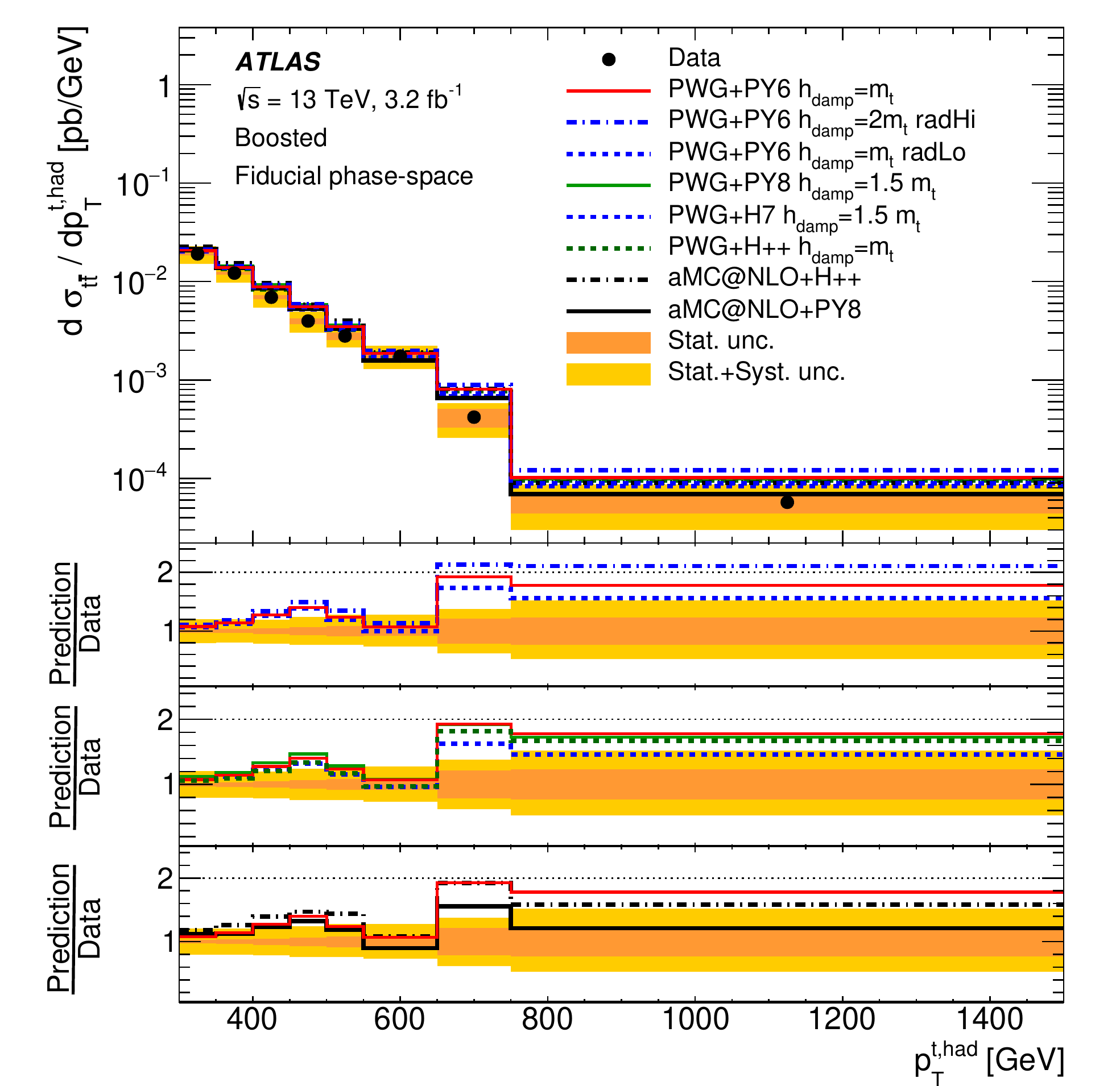}
}
\subfloat[\label{fig:ATLAS:diffxs:absrap:ljets:13tev}]
{
\includegraphics[width=0.32\linewidth]{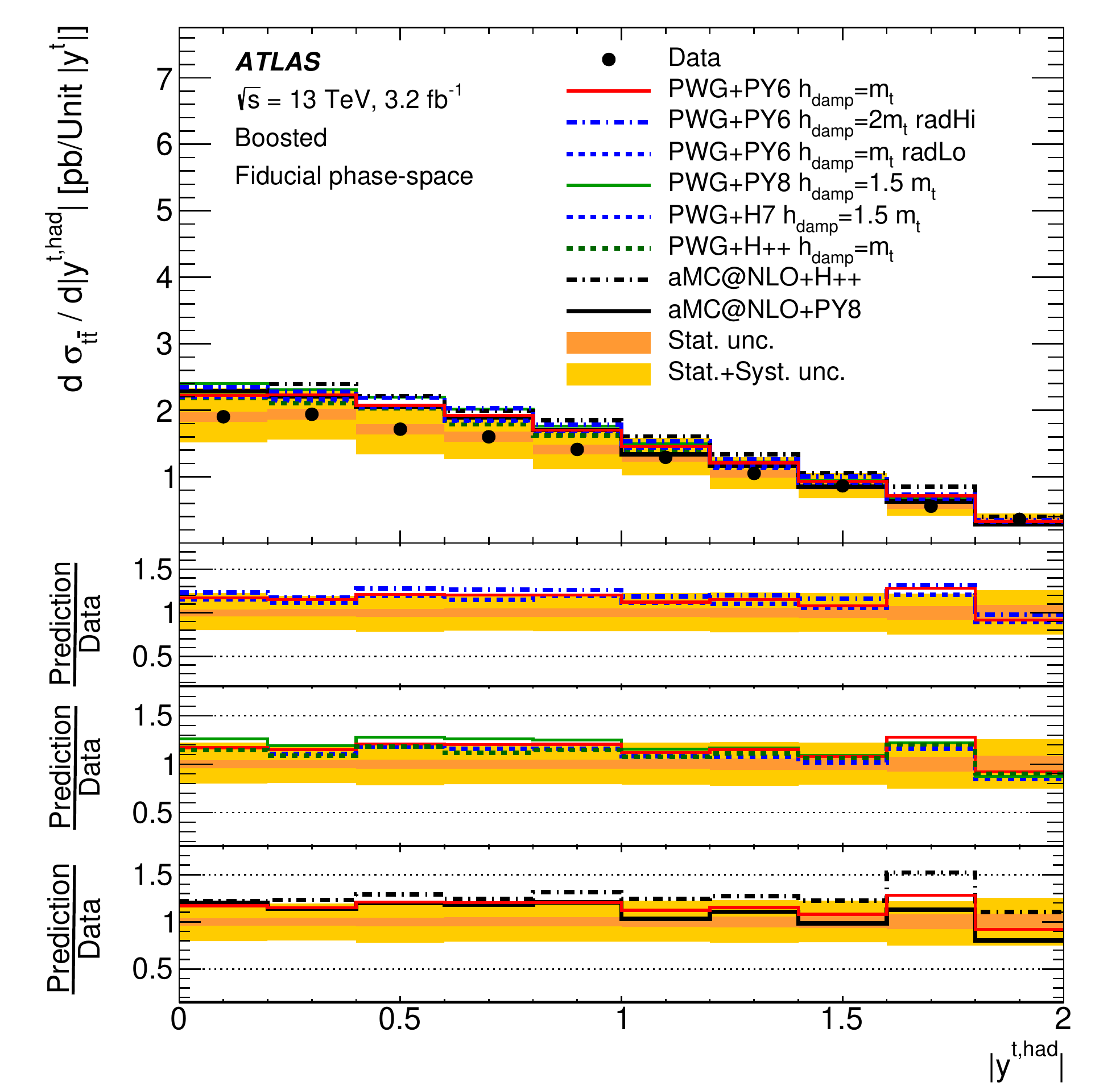}
}
\subfloat[\label{fig:ATLAS:diffxs:pt2:fullhad:13tev}]
{
\includegraphics[width=0.32\linewidth]{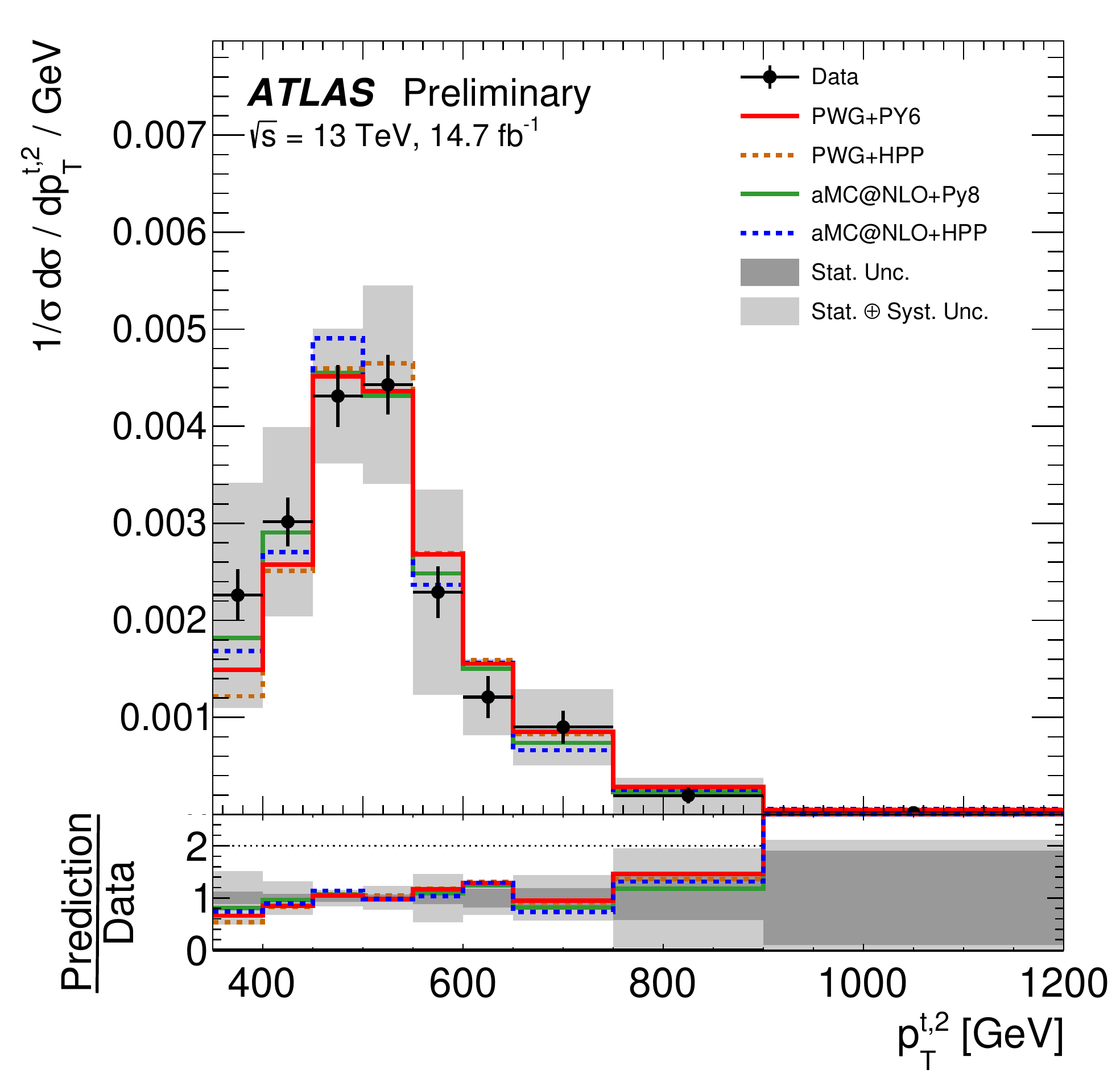}
}
\caption{Fiducial phase-space differential cross sections as a function of \protect\subref{fig:ATLAS:diffxs:pt:ljets:13tev} the  transverse momentum ($\pt^{\thad}$)
and  \protect\subref{fig:ATLAS:diffxs:absrap:ljets:13tev} the absolute value of the rapidity ($\left|y^{\thad}\right|$) of the hadronic top quark measured by the ATLAS experiment at $\sqrt{s}=13\tev$ in the lepton+jets channel \cite{ATLAS:xs:ljets:13tev}.
\protect\subref{fig:ATLAS:diffxs:pt2:fullhad:13tev} normalized fiducial phase-space differential cross sections as a function of the transverse momentum of the second-leading top-quark jet measured by the ATLAS experiment at $\sqrt{s}=13\tev$ in the full hadronic channel \cite{ATLAS:xs:fullhad:13tev}.
\label{fig:ATLAS:tt:xs:13tev}}
\end{figure}

In Figure~\subref*{fig:ATLAS:diffxs:pt:ljets:13tev} and Figure~\subref*{fig:ATLAS:diffxs:absrap:ljets:13tev} are presented the measured absolute differential cross sections as a function of the transverse momentum and absolute rapidity of the hadronically-decaying top quark. The results are in fair agreement with the predictions over a wide kinematic range for the different observables. However, most generators predict an higher cross section at high values of $\pt^{\thad}$ compared to the
observations in data, as it has already been shown by the different Run 1 measurements by both ATLAS and CMS at 8 TeV \cite{ATLAS:xs:ljets:8tev,CMS:xs:ljets:8tev}. 

\subsection{Top pair differential cross sections in the full hadronic channel
at 13 TeV}

The measurement of $t\bar{t}$ differential cross sections in the full hadronic channel has been performed by both ATLAS \cite{ATLAS:xs:fullhad:13tev} and CMS \cite{CMS:xs:fullhad:13tev} at
a centre-of-mass energy $\sqrt{s}=13\tev$. The integrated luminosities for these analyses correspond to $L = 14.7 \invfb$
(recorded in 2015 and 2016 in proton-proton collisions) and $L=2.53\invfb$ (recorded in 2015), respectively.

Boosted top quark events in full hadronic channels are selected by requiring two high-$p_{T}$ top-tagged jets, according the procedures described in  \cite{ATLAS:boosted:reco:13tev,CMS:boosted:reco:13tev}.

For these analyses, the main challenge is represented by the estimation of the multijet background. Both measurements employ data-driven methods to estimate this background:
\begin{itemize}
\item the ATLAS measurement makes use of an extension of the ABCD method to extract the yield in the signal region based on the yields in control regions using the formula
\begin{equation}
S=\frac{1}{2}\left(\frac{G}{A}+\frac{H}{B}\right)C \mathrm{,}
\end{equation}
where the control regions $A,B,C,G,H$ are defined in terms of the number of $b$- and $t$-tagged jets;
\item the CMS measurement makes use of a template fit of the distribution of the soft-drop groomed jet mass to extract both the signal and multijet event yields. The signal and QCD templates
are taken from the simulation and the control sample in data, respectively, corrected with a
MC transition factor.
\end{itemize}
%

ATLAS  measured the normalized differential cross section in the fiducial phase space as a function of the $\pt$ and absolute rapidity of the leading and subleading top, the $\pt$, mass and absolute rapidity of the top-antitop system and additional variables sensitive to effects of initial- and final-state radiation, to the different parton
distribution functions (PDF), and to non-resonant processes including particles beyond the SM. The differential cross section as a function of the $\pt$ of the subleading top is shown in Figure~\subref*{fig:ATLAS:diffxs:pt2:fullhad:13tev}.

\begin{figure}[htbp]
\centering
\subfloat[\label{fig:CMS:diffxs:detector:fullhad:13tev}]
{
\includegraphics[width=0.34\linewidth]{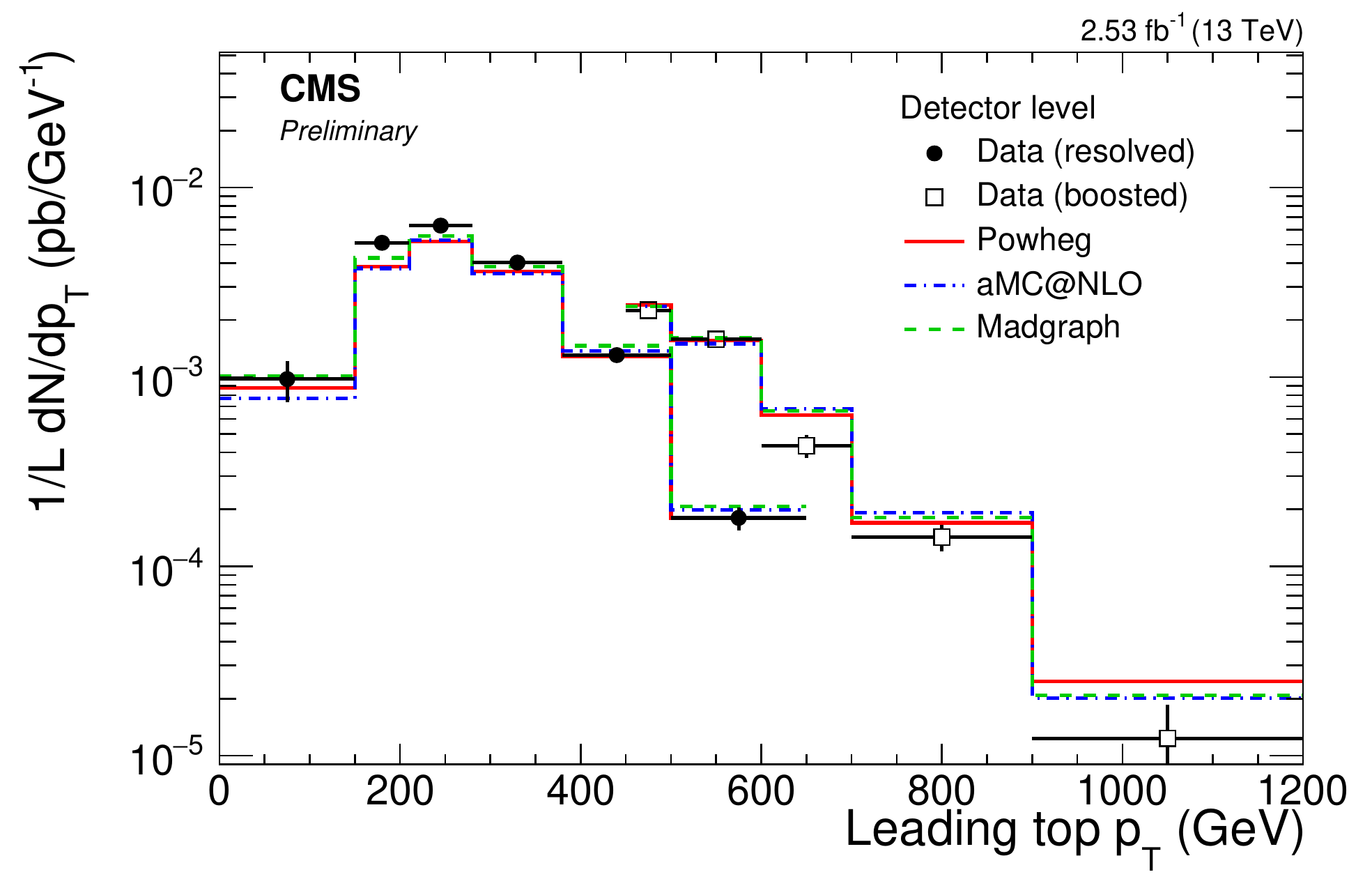}
}
\subfloat[\label{fig:CMS:diffxs:parton:fullhad:13tev}]
{
\includegraphics[width=0.34\linewidth]{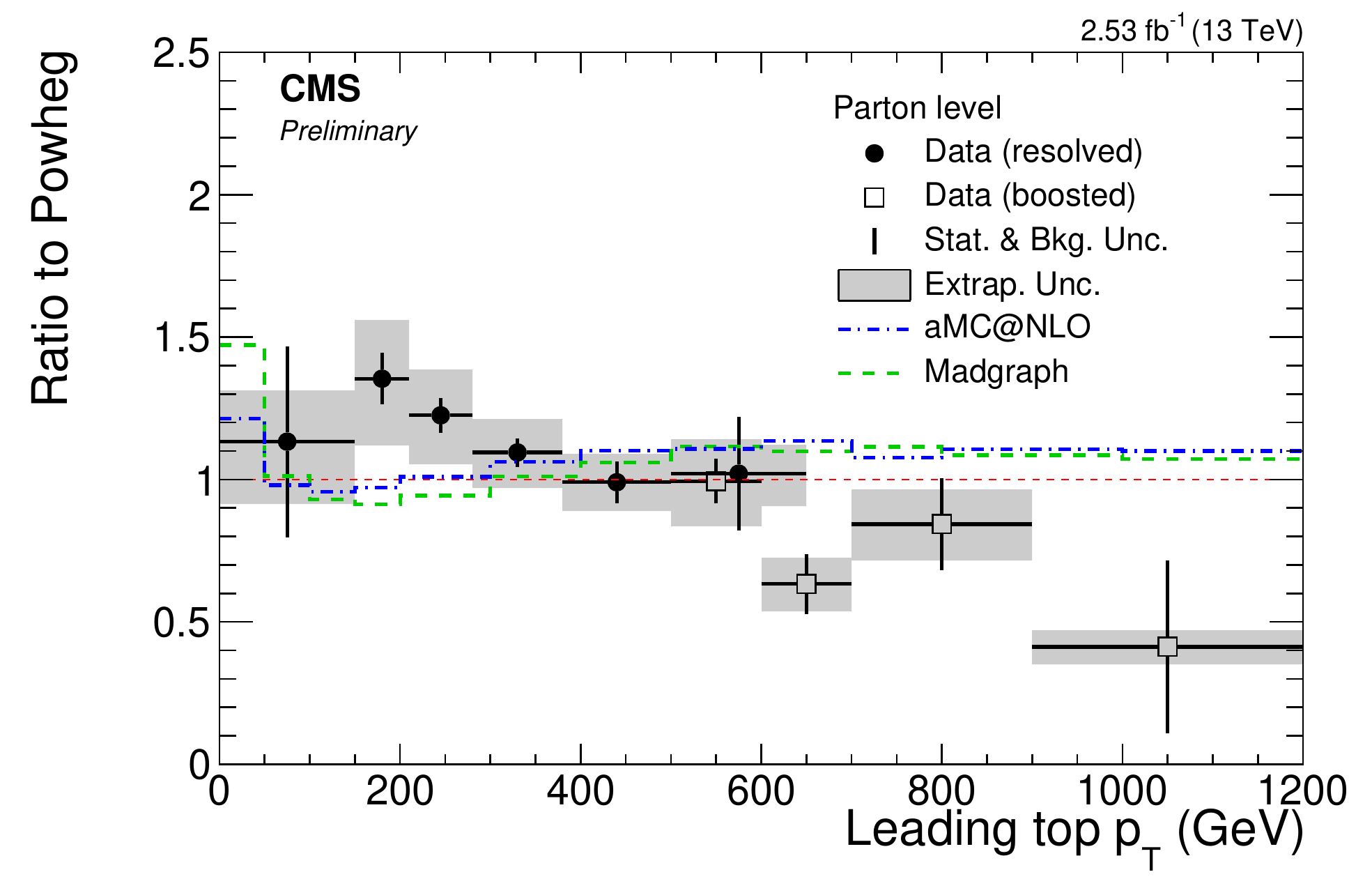}
}
\subfloat[\label{fig:CMS:jetmass:abs}]
{
\includegraphics[width=0.28\linewidth]{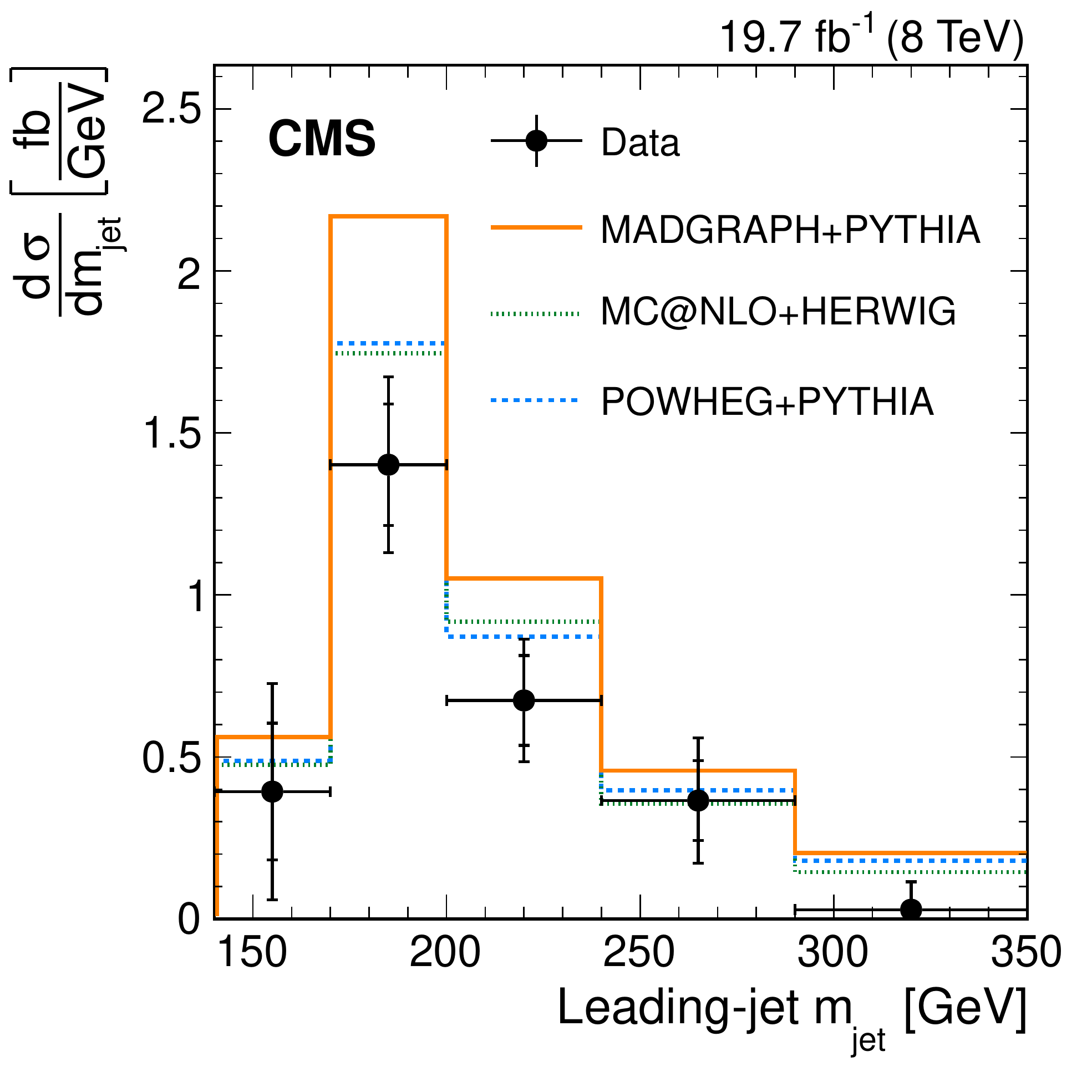}
}
\caption{\protect\subref{fig:CMS:diffxs:detector:fullhad:13tev} detector-level differential cross section as a function of the transverse momentum of the leading top measured by the CMS expertiment at $\sqrt{s}=13\tev$ in the full hadronic channel using resolved (black dots) and boosted (empty boxes) techniques \cite{CMS:xs:fullhad:13tev}. \protect\subref{fig:CMS:diffxs:parton:fullhad:13tev} ratio between the parton-level differential cross section as a function of the transverse momentum of the leading top measured by the CMS expertiment at $\sqrt{s}=13\tev$ in the full hadronic channel using resolved (black dots) and boosted (empty boxes) techniques and the prediction obtained with the POWHEG+PYTHIA event generator \cite{CMS:xs:fullhad:13tev}. \protect\subref{fig:CMS:jetmass:abs} particle-level absolute  differential cross section as a function of the leading jet mass measured by the CMS experiment at $\sqrt{s}=8\tev$ in the lepton+jets channel \cite{CMS:xs:jetmass}.\label{fig:CMS:tt:xs:13tev}}
\end{figure}

CMS  measured the inclusive cross section, detector and parton level differential cross sections as a function of the $\pt$ of the leading top quark, shown in Figure~\subref*{fig:CMS:diffxs:detector:fullhad:13tev} and Figure~\subref*{fig:CMS:diffxs:parton:fullhad:13tev}. The~ measured inclusive cross section is
\begin{displaymath}
\sigma_{t\bar{t}} = 727 \pm 46(\mathrm{stat.})^{+115}_{-112}(\mathrm{syst.}) \pm 20(\mathrm{lumi.})\, \mathrm{pb}\mathrm{;}
\end{displaymath}
in agreement, within the uncertainties, with the theoretical cross section $\sigma_\ttbar^{th} = 832^{+20}_{-29} (\mathrm{scale}) \pm 35 (\mathrm{PDF} + \alpha_s)\,\pb$, as calculated
with the TOP++ program \cite{toppp} at next-to-next-to-leading order (NNLO) in perturbative
QCD, including soft-gluon resummation at next-to-next-to-leading-log order. 
%

\subsection{Measurement of the jet mass distribution in boosted top pair
production at 8 TeV}
A detailed understanding of jet substructure observables, most
importantly the jet mass $m_\mathrm{jet}$, is crucial for LHC analyses in boosted topologies. 
CMS has performed, for the first time, a measurement of the differential $\ttbar$ production cross section as
a function of the leading jet mass in fully-merged top quark decays \cite{CMS:xs:jetmass}. Data collected
with the CMS detector in $pp$ collisions at $\sqrt{s} = 8\tev$ are used, corresponding to an
integrated luminosity of $19.7\invfb$. The measurement is carried out in the lepton+jets channel. 

The products of the hadronic decay are reconstructed with a single
Cambridge/Aachen jet with distance parameter $R = 1.2$, and transverse momentum
$\pt > 400\gev$. No top tagging algorithm is applied to the large-$R$ jet, in order to avoid any bias to the $m_\mathrm{jet}$ distribution.
A bias would for example be introduced by selecting the leading jet based on the number of
subjets or requiring a certain maximum value of the N-subjettiness as applied in top-tagging
algorithms.

The measured absolute differential cross section is shown in Figure~\subref*{fig:CMS:jetmass:abs}. The measurement gives a total cross~ section of $103.5 \pm 11.2 \left(\mathrm{stat.}\right)\pm 10.9\left(\mathrm{syst.}\right)\pm 9.4\left(\mathrm{model}\right)\fb$. The predicted cross sections from the MADGRAPH+PYTHIA and POWHEG+PYTHIA $\ttbar$ simulations, normalized to a total $\ttbar$ cross section of 252.9 pb, are 160 fb and 134 fb respectively.

%

The peak position of the $m_\mathrm{jet}$ distribution is also sensitive to the top
quark mass $m_t$. The value of $m_t$ is determined
using the normalized measurement since only the shape of the $m_\mathrm{jet}$ distribution can be reliably
calculated.  The measured value of $m_t$ is extracted from data using a fit of the $\chi^2$ evaluated comparing the mesaured $m_\mathrm{jet}$ distribution with templates obtained from MADGRAPH+PYTHIA predictions with different values of $m_t$. 
The result is
\begin{displaymath}
m_t  = 171.8 \pm 5.4  \left(\mathrm{stat.}\right) \pm 3.0 \left(\mathrm{syst.}\right) \pm  5.5  \left(\mathrm{model}\right) \pm 4.6 \left(\mathrm{theory}\right)\gev = 171.8 \pm 9.5 \gev\mathrm{,}
\end{displaymath}
consistent with the current LHC+Tevatron average, $173.34 \pm 0.27 \left(\mathrm{stat.}\right) \pm 0.71 \left(\mathrm{syst.}\right)\gev$ \cite{mass:comb}.

%% file: charge_asymmetry_short.tex
\section{Charge asymmetry measurement in highly boosted top pair production at 8 TeV}
In the $pp\rightarrow\ttbar$ process the angular distributions of top and anti-top quarks are expected
to present a subtle difference, which could be enhanced by processes not included in the
Standard Model. In the SM, a forward-backward asymmetry ($A_\mathrm{FB}$), of order $\alpha_s$, is expected at a proton-antiproton ($p\bar{p}$) collider such as the Tevatron. 

At the LHC, the forward-backward asymmetry is not present due to the symmetric
initial state, but a related charge asymmetry  ($A_\mathrm{C}$) is expected in the distribution of the difference of
absolute rapidities of the top and anti-top quarks:
\begin{equation}
A_\mathrm{C}=\frac{N\left(\Delta\left|y\right|>0\right)-N\left(\Delta\left|y\right|<0\right)}{N\left(\Delta\left|y\right|>0\right)+N\left(\Delta\left|y\right|<0\right)}\mathrm{,}
\end{equation}
where $\Delta\left|y\right| = \left|y_t\right| - \left|y_{\bar{t}}\right|$ and $y$ denotes the rapidity of the top and anti-top quarks.

The ATLAS experiment has performed a  measurement of the rapidity-dependent charge asymmetry for boosted top-quark pair production 
in the lepton+jets channel \cite{ATLAS:chargeasymm} using the boosted top reconstruction and tagging techniques described in \cite{ATLAS:boosted:reco:8tev}.

The analysis is performed using $20.3 \invfb$ of $pp$ collision data at $\sqrt{s} = 8\tev$ collected by the ATLAS experiment at the
LHC. The charge asymmetry in a fiducial region with large invariant mass of
the top-quark pair ($m_\ttbar > 0.75 \tev$) and an absolute rapidity difference of the top and antitop
quark candidates within $-2 < \left|y_t\right| - \left|y_{\bar{t}}\right| < 2$ is measured to be $A_\mathrm{C}=\left(4.2 \pm 3.2\right)\%$, in agreement
with the Standard Model prediction at next-to-leading order $A_\mathrm{C}^{th}=\left(1.60\pm0.04\right)\%$ \cite{chargeasymm:th}.

A differential measurement has been also performed in
three $\ttbar$ mass bins ($0.75\tev < m_\ttbar < 0.9\tev$, $0.9\tev<m_\ttbar<1.3\tev$ and $m_\ttbar>1.3\tev$). The most significant deviation from the
SM prediction (around $1.6\sigma$) is observed in the mass bin that ranges from 0.9 TeV to 1.3 TeV. The other two mass bins yield values compatible with the SM prediction within $1\sigma$.

%% file: summary.tex
\section{Summary and outlook}
LHC offers a unique opportunity to explore extreme topologies through boosted tops. Several boosted top reconstruction algorithms have
been (and are being) developed by ATLAS and CMS: jet grooming procedures allow for stability
in high pileup conditions  and the use of substructure variables improves the background
discrimination.

Boosted tops have been used in SM measurements and BSM searches (not presented in this paper): ATLAS and CMS have measured the $\ttbar$ differential cross sections using boosted tops in both the lepton+jets and full hadronic channel and the charge asymmetry in the lepton+jets channel. We are entering an era where data statistics is not the limiting factor: the measurements are already limited by uncertainties related to the large-$R$ jets and signal modelling.

The measurements presented in this paper 
can be used to improve future analyses:
\begin{itemize}
\item by adding better constraints to the generator parameters;
\item by improving the understanding of the jet mass distribution.
\end{itemize}

In the light of the ongoing data taking at the LHC at $\sqrt{s} = 13 \tev$, boosted
techniques will gain even more importance for measurements of top quarks with high transverse momenta but also for  searches for new phenomena at the TeV scale. The experience gained
with boosted techniques in Run-1 and early Run-2 analyses is a promising basis for the upcoming challenges.